\documentclass{amsart}
\usepackage{amssymb}
\usepackage{amsfonts}
\usepackage{amsmath}
\usepackage[legalpaper,bookmarks=true,colorlinks=true,linkcolor=blue,citecolor=blue]{hyperref}
\usepackage{graphicx}%
\usepackage{fancyhdr}
\usepackage{color}
\usepackage[mathlines]{lineno}
\usepackage{lscape}
\usepackage{epsfig}
\usepackage{natbib}
\usepackage{geometry}
\usepackage{tgbonum}
\usepackage[]{listings}

\fontfamily{qcr}\selectfont

\newtheorem{theorem}{Theorem}
\theoremstyle{plain}

\newtheorem{proposition}{Proposition}

\numberwithin{equation}{section}

\newcommand{\Bin}{\bigskip \noindent}

\newcommand{\Ni}{\noindent}

\begin{document}
\Large
\title[Extreme Value Theory and China Car insurance]{Applying of the Extreme Value Theory for determining extreme claims
in the automobile insurance sector: Case of a China car insurance}

\author{Daouda Diawara}

\author{Ladji Kane}

\author{Soumaila Dembele}

\author{Gane Samb Lo}


\begin{abstract} According to the Chinese Health Statistics Yearbook, in 2005, the number
of traffic accidents was 187781 with total direct property losses of
103691.7 (10000 Yuan). This research aims to fill the gap in the literature
by investigating the extreme claim sizes not only for the entire portfolio.
This empirical study investigates the behavior of the upper tail of the
claim size by class of policyholders.\\

\Ni $^{\dag}$ Daouda Diawara\\
Universit\'e des Sciences Sociale et de Gestion de Bamako ( USSGB)\\
Facult\'e des Sciences \'Economiques et de Gestion (FSEG) Bamako, Mali\\
Daouda Diawara: btddiawara@yahoo.fr\\

\Ni $^{\dag\dag}$ Ladji Kane\\
Universit\'e des Sciences Sociale et de Gestion de Bamako ( USSGB)\\
Facult\'e des Sciences \'Economiques et de Gestion (FSEG) Bamako, Mali\\
fsegmath@gmail.com\\

\noindent $^{\dag\dag\dag}$ Soumaila Dembel\'e\\
Universit\'e des Sciences Sociale et de Gestion de Bamako ( USSGB)\\
Facult\'e des Sciences \'Economiques et de Gestion (FSEG)\\
Email: dembele.soumaila@ugb.edu.sn, soumailadembeleussgb@gmail.com\\

\noindent $^{\dag\dag\dag\dag}$ Gane Samb Lo.\\
LERSTAD, Gaston Berger University, Saint-Louis, S\'en\'egal (main affiliation).\newline
LSTA, Pierre and Marie Curie University, Paris VI, France.\newline
AUST - African University of Sciences and Technology, Abuja, Nigeria\\
gane-samb.lo@edu.ugb.sn, gslo@aust.edu.ng, ganesamblo@ganesamblo.net\\
Permanent address : 1178 Evanston Dr NW T3P 0J9,Calgary, Alberta, Canada.\\



\end{abstract}
\maketitle

\newpage

\section{Introduction} \label{sec1}

\subsection{Context of the study}

\noindent China has been experiencing a high-speed urbanization and
motorization related to rapid economic growth. The growing urbanization and
motorization levels result in increasing frequencies of road traffic
accidents, injuries and deaths. As reported by the world Health
Organization, in 2015, the total fatalities in china are among the highest
in the world. According to the Chinese Health Statistics Yearbook, in 2005,
the number of traffic accidents was 187781 with total direct property losses
of 103691.7 (10000 Yuan). The Chinese car insurance and reinsurance
companies mainly support these losses. Moreover, statistics highlight that
322 events are considered as serious accidents and generate very large
losses for the insurers. The appearance of these excessively large claims
leads to think about claims resulting from severs accidents as with massive
pile up of vehicles or light-truck accidents. In other instances, an insurer
might be confronted with large claims coming from a policy involving very
valuable items such as concentrated risks (e.g. a large number of luxury
cars located in the same area).\newline

\noindent This research is a contribution to the statistical investigating of the claims. We do not focus, as in many studies, on the entire portfolio. Rather, we will investigate the behavior of the upper tail of the claim size by class of policyholders.\newline

\Ni Since we use the univariate extreme value theory, we feel obliged to give a brief account of such a theory 

\subsection{An easy introduction of the univariate extreme value theory}.

\Ni The theory is usually presented in tow main point of views: the max-stability apporach (MSA) and the Peak Over Threshold (POT) Method.\\

\subsubsection{The max-stability approach}

\Ni Let $X$, $X_1$, $X_2$, $\cdots$ be a sequence independent real-valued randoms, defined on the same probability space $(\Omega, \mathcal{A},\mathbb{P})$, with common cumulative distribution function $F$, which has the lower and upper endpoints,  \index{lower and upper endpoint} the first asymptotic moment  \index{asymptotic moment} function and the generalized inverse function  \index{generalized inverse function} respectively defined by

$$
lep(F)=\inf \{x \in \mathbb{R}, \ F(x)>0\}, \  uep(F)=\sup \{x \in \mathbb{R}, \ F(x)<1\} 
$$

$$
R(x,F)=\frac{1}{1-F(x)}\int_{x}^{uep(F)}(1-F(y)) \ dy, \ x \in ]lep(F),uep(F)[
$$

\Bin and

$$
F^{-1}(u)=\inf \{x \in \mathbb{R}, \ F(x)\geq y\} \ for \ u \in ]0,1[ \ and \ F^{-1}(0)=F^{-1}(0+).
$$

\Bin The main task the \textit{UEVT}, originally, was the study of the max-stability of iid sequence. Namely, the essential problem is finding
real and nonrandom sequences \ $(a_{n}>0)_{n\geq 1}$ and $(b_{n})_{n\geq 1}$ such that the centered and normalized sequence of partial maxima
$$
\frac{\max(X_1,\cdots,X_n)-b_{n}}{a_{n}}=:\frac{X_{n,n}-b_{n}}{a_{n}}
$$

\Bin (with $X_{n,n}=\max(X_1,\cdots,X_n))$ weakly converges to some \textit{cdf} $M$, which is equivalent to:  for any continuity point $x$ of the \textit{cdf} $G$ of $M$:

\begin{equation}
\lim_{n\rightarrow \infty} \mathbb{P}\left(\frac{X_{n,n}-b_{n}}{a_{n}}\leq
x\right)=\lim_{n\rightarrow \infty }F^{n}(a_{n}x+b_{n})=M(x).  \label{dl05}
\end{equation}

\Bin The \textit{UVT} goes back to the earlier years of the 1900's with many contributors as \cite{fishertippet}, \cite{gumbel}, etc. for statistical purposes. On the mathematical ground, \cite{gnedenko1943} (see \cite{ips-wciia-ang} [Theorem 2, page 13] and \cite{resnick}) have final  characterizations of max-stable random variables as stated in Formula \eqref{GEV} below. First important accounts of the theory are given in \cite{galambos}, \cite{dehaan}, etc. The \textit{UEVT} uses regularly varying functions, a concept deeply described in \cite{feller2} (page 275) and \cite{loeve} (page 354). Later several authors, on the basis of these fundamental texts, provided other exposition of the theory, some of them focusing on what is now called \textit{statistics of extremes}. Let use cite a few of them \cite{dehaan2} (who significantly developed his pioneering work in \cite{dehaan}), \cite{resnick}, \cite{Embrechts}, \cite{bgt}, \cite{ips-wciia-ang}. Other important references can be found in the cited books.

\Bin It is showed that the convergence in \eqref{dl05} is a convergence in type, meaning that a change of the coefficients $(a_{n}>0)_{n\geq 1}$ and $(b_{n})_{n\geq 1}$
to $(\alpha_{n}>0)_{n\geq 1}$ and $(\beta_{n})_{n\geq 1}$ and a change of $M$ to $M_1$ with \textit{cdf} $G_1$ necessarily leads to the following facts:

\begin{equation}
\alpha_n/a_n \rightarrow A>0 \ \ and \ \ (\beta_n-\alpha_n)/a_n \rightarrow 0 \ \  as \ \ n\rightarrow +\infty \label{changeA}
\end{equation}

\Bin and

\begin{equation}
\forall x\in \mathbb{R}, \ G_1(x)=G(Ax+B), \label{changeB}
\end{equation}

\Bin (see \cite{ips-wciia-ang} [Lemma 42, page 12] and \cite{resnick}) The later fact says that $G_1$ and $G$ are equal in type. Gnedenko's theorem establishes that if \eqref{dl05} holds and if $M$ is not concentrated on a single point, the only three possibilities are the following:

\begin{equation}
Fr_{\alpha}(x)=\exp(-x^{1/\gamma}) \ 1_{(x\geq 0)}, \ \  (Type \ I) \label{limitsI}
\end{equation}

\begin{equation}
W_{\alpha}(x)=\exp(-(-x)^{1/\gamma}) \ 1_{(x\geq 1)}+1_{(x>0)}, \ \ (Type \ II) \label{limitsII}
\end{equation}

\Bin and

\begin{equation}
\Lambda(x)=\exp(-e^{-x}) \ 1_{\mathbb{R}}(x), \ \ (Type \ III) \label{limitsIII}
\end{equation}

\Bin Let us apply Formulas \eqref{changeA} and \eqref{changeB} to \eqref{limitsI} with $A=\gamma=\alpha$ and $B=1$, next Formulas \eqref{changeA} and \eqref{changeB} to \eqref{limitsI}  with $A=-\gamma=-1/\beta$ and $B=-1$ and finally by interpreting 

$$
\biggr[\exp(-(1+\gamma x)^{-1/\gamma})\biggr]_{\gamma=0}=\exp\left(-\exp(-x)\right), \ x \in \mathbb{R}.
$$

\Ni to get that any weak $M$ in \eqref{dl05} has a \textit{cdf} in the family of the Generalized Extreme
Value (GEV) \textit{df} : 

\begin{equation}
H_{\gamma }(x)=\exp (-(1+\gamma x)^{-1/\gamma })\text{, }1+\gamma x\geq 0, \label{GEV}
\end{equation}

\Bin parametrized by $\gamma \in \mathbb{R}$, with $H_{0}(x)=1-\exp(-e^{-x})$, $x \in \mathbb{R}$, for $\gamma=0$. The parameter $\gamma $ is called
the extreme value index.\\

\Bin As to the choice of the sequences $(a_n)_{n\geq 1}$ and $(b_n)_{n\geq 1}$, the following ones

$$
a_n=F^{-1}\left(1-\frac{1}{n}\right) \ \ and \ \ b_n=0, 
$$

$$
a_n=F^{-1}\left(1-\frac{1}{n}\right)-uep(F) \ \ and \ \ b_n=-uep(F),
$$

\Bin (with $uep(F)<+\infty$ necessarily) and

$$
a_n=F^{-1}\left(1-\frac{1}{ne}\right)-F^{-1}\left(1-\frac{1}{n}\right) \ \ and \ \ b_n=F^{-1}\left(1-\frac{1}{n}\right),
$$

\Bin for $n\geq 1$, leads to the the limits \eqref{limitsI}, \eqref{limitsII} and \eqref{limitsIII} respectively.\\
 
\Ni Now, we are recalls useful criteria for \textit{cdf}'s to belong to  the whole domain 

$$
\mathcal{D}=\{G_\gamma, \ \gamma>0\}
$$ 

\Bin of extreme attraction and functional representation of \textit{cdf}'s and their quantile functions in $\mathcal{D}$.\\

\Bin First, we have: 

\begin{proposition} \label{prop1} \label{evtCriteria} (Main criteria for $F \in \mathcal{D}$)

\Ni Let $F$ be a \textit{cdf} on $\mathbb{R}$. We have two general cases and in each of them, we consider two sub-cases.\\

\noindent \textbf{(A)} - Let $uep(F)=+\infty$.\\

\noindent \textbf{(A1) - $F \in D(G_{1/\gamma})$}, $\gamma>0$, if one of the following assertions hold.

$$
\forall \lambda>0, \ \lim_{x\rightarrow +\infty} \frac{1-F(\lambda x)}{1-F(x)}=\lambda^{-1/\gamma}. \ \ \ (A11)
$$

$$
\forall \lambda>0, \lim_{u\rightarrow 0} \frac{F^{-1}(1-\lambda u)}{F^{-1}(1-u)}=\lambda^{-\gamma}. \ \ \ (A12)
$$

$$
\lim_{x\rightarrow +\infty} \frac{x\ F^{\prime}(x)}{1-F(x)}=1/\gamma. \ \ \ (A13)
$$

\Bin \textbf{(A2) - $F \in D(G_{0})$} if one of the following assertions hold.

\Bin  (a) There exists a slowly varying function  \index{slowly varying function} at zero that we denote as $s(u)$ of $u\in ]0,1[$ such that

$$
\forall \lambda>0, \ \lim_{u\rightarrow 0} \frac{F^{-1}(1-\lambda u)-F^{-1}(1-u)}{s(u)}=- \log \lambda. \ \ \ (A21)
$$

\Bin (b) The $\Gamma$-variation formula holds (due to \cite{dehaan}).

$$
\forall t \in \mathbb{R},  \lim_{x\rightarrow uep(F)} \frac{1-F(x+tR(x)}{1-F(x)}=e^{-t} \ \ \ (A22)
$$

\Bin (c) Upon the twice differentiability of $F$ on the left neighborhood of $uep(F)$, the following Von-Mises condition holds:

$$
\lim_{x\rightarrow +\infty} \frac{F^{\prime\prime}(x) F(x)}{(F^{\prime}(x))^2}=-1. \ \ \ (A23)
$$

\Bin (d) Upon the twice differentiability of $F$ on the left neighborhood of $uep(F)$, the function (\cite{lo86}'s criteria)

$$
s(u)=-u\left(F^{-1}(1-u)\right)^{\prime} 
$$

\bigskip \noindent is slowly varying at zero.\\

\Bin \textbf{(B)} - Let $uep(F)<+\infty$.\\

\Bin \textbf{(B1). $F \in D(G_{\gamma})$}, $\gamma<0$ if one of the following assertions holds.

$$
\forall \lambda>0, \ \lim_{x\rightarrow uep(F)} \frac{1-F(uep(F)- (\lambda x)^{-1})}{1-F(uep(F)-x^{-1})}=\lambda^{1/\gamma}. \ \ \ (B11)
$$

$$
\forall \lambda>0, \lim_{u\rightarrow 0} \frac{uep(F)-F^{-1}(1-\lambda u)}{uep(F)-F^{-1}(1-u)}=\lambda^{-\gamma}. \ \ \ (B12)
$$

$$
\lim_{x\rightarrow uep(F)} \frac{(upe(F)-x)\ F^{\prime}(x)}{1-F(x)}=-1/\gamma. \ \ \ (B13)
$$

\Bin  \textbf{(B2)} To test whether or not $F \in D(G_{0})$, we re-use the criteria of Sub-case (A2).

\end{proposition}

\bigskip \noindent Next, we have the main representations of $F^{-1}$ for $F \in \mathcal{D}$.

\begin{proposition} \label{portal.rd} (Representations of quqntile functions whithin the extreme attraction domain) We have the following characterizations for the three extremal domains.

\bigskip \noindent (a) $F\in D(H_{\gamma })$, $\gamma >0$, if and only if there exist a constant $c$ and functions $a(u)$ and $\ell (u)$ of $u\rightarrow $ $u\in ]0,1]$ satisfying

\begin{equation*}
(a(u),\ell (u))\rightarrow (0,0)\text{ as }u\rightarrow +\infty ,
\end{equation*}

\Bin such that $F^{-1}$ admits the following representation of \cite{karamata} \index{karamata} \index{representation of Karamata}

\begin{equation}
F^{-1}(1-u)=c(1+a(u))u^{-\gamma }\exp \left(\int_{u}^{1}\frac{\ell (t)}{t}dt\right). \label{portal.rdf}
\end{equation}

\bigskip \noindent (b) $F\in D(H_{\gamma })$, $\gamma <0,$ if and only if $uep(F)<+\infty $ and there exist a constant $c$ and functions $a(u)$ and $\ell (u)$ of $u\in ]0,1]$ satisfying

\begin{equation*}
(a(u),\ell (u))\rightarrow (0,0)\text{ as }u\rightarrow +\infty ,
\end{equation*}

\bigskip \noindent such that $F^{-1}$ admit the following representation of \cite{karamata} \index{representation of Karamata}

\begin{equation}
uep(F)-F^{-1}(1-u)=c(1+a(u))u^{-\gamma }\exp \left(\int_{u}^{1}\frac{\ell (t)}{t} dt\right). \label{portal.rdw}
\end{equation}

\bigskip \noindent (c) $F\in D(H_{0})$ if and only if there exist a constant $d$ and a slowly varying function  \index{slowly varying function} $s(u)$ such that

\begin{equation}
F^{-1}(1-u)=d+s(u)+\int_{u}^{1}\frac{s(t)}{t}dt,0<u<1, \label{portal.rdg}
\end{equation}

\bigskip \noindent and there exist a constant $c$ and functions $a(u)$ and $\ell (u)$ of $ u\rightarrow $ $u\in ]0,1]$ satisfying

\begin{equation*}
(a(u),\ell (u))\rightarrow (0,0)\text{ as }u\rightarrow +\infty ,
\end{equation*}

\bigskip \noindent such that $s$ admits the \cite{dehaan} representation

\begin{equation}
s(u)=c(1+a(u)) \exp \left(\int_{u}^{1}\frac{\ell (t)}{t}dt\right). \label{portal.rdgs}
\end{equation}

\Bin Moreover, if $F^{-1}(1-u)$ is differentiable for small values of $s$ such that $r(s)=-s(F^{-1}(1-s))^{\prime }=udF^{-1}(1-s)/ds$ is slowly varying at zero, then \ref{portal.rdg} may be replaced by

\begin{equation}
F^{-1}(1-u)=d+\int_{u}^{u_{0}}\frac{r(t)}{t}dt, \ 0<u<u_{0}<1, \label{portal.rdgr}
\end{equation}

\Bin which will be called a \cite{lo86}'s representation or a\textit{reduced \cite{dehaan} \index{de haan} representation} of $F^{-1}.$

\end{proposition}

\Bin \textbf{Remark}. Let us remark that any $F \in \mathcal{D}$ is associated to a paire of functions $(a(u),b(u))$ of $u \in ]0,1[$. For $\gamma\neq 0$, these functions directly appear in the representations \eqref{portal.rdf} and \eqref{portal.rdw} in Proposition \label{portal.rd}. For $\gamma=0$, the representation uses $s(\circ)$  which, in turn, uses the function $(a(u),b(u))$ of $u \in ]0,1[$. In that sense each $F \in \mathcal{D}$ can be represented as
$F\equiv (a,b)$.\\

\Ni Let us finish by stating a rule for differentiable \textit{cds}'s.\\

\Bin \textbf{Rules of working \label{commentsII}}. In the domain of extremal attraction, most of the \textit{cdf}'s which are used in applications are differentiable in a left-neighborhood of the upper endpoint. In such a case, we may take $a\equiv0$ in Representation \eqref{portal.rdf} and \eqref{portal.rdw} in Proposition \ref{portal.rd}. In that case, we need only the function $b(\circ)$ and by solving easy differential equations, we may take

\begin{equation}
b(u)=-u(G^{-1}(1-u))^{\prime }-\gamma, \ u\in (0,1) \ \ and \ \ a\equiv 0
\end{equation}

\Bin  for $\gamma>0$ and

\begin{equation}
b(u)= -\gamma -\frac{u}{F^{\prime}\biggr(F^{-1}(1-u)\biggr)\biggr(uep(F)-F^{-1}(1-u)\biggr)}, \ u\in ]0,1[.
\end{equation}

\Bin for $\gamma<0$, whenever we have  $b(u)\rightarrow 0$ as $u\rightarrow 0$. Consequently, we may drop $a_n$ in the rates of convergence to reduce them to $O_{\mathbb{P}}(b_n \vee c_n)$.\\

\Bin For $\gamma=0$, Representation \eqref{portal.rdgr} in Proposition \ref{portal.rd} holds for

$$ 
s(u)=-u(F^{-1}(1-u))^{\prime }, \ 0<u<1,
$$

\Bin whenever it is slowly varying at zero and the rate of convergence $a_n$ becomes useless. In such cases, the rate of convergence reduces $O_{\mathbb{P}}(d_n \vee c_n)$.\\

\subsubsection{The POT approach}
That approach relies on the Generalized Pareto Distribution (GDP) with two parameters $\lambda \in \mathbb{R}$ and $b>0$, defined as follow

$$
G_{\lambda,b}(x)=\biggr(1-(1+\lambda x/b)^{-1/\lambda}\biggr) 1_{(\lambda\neq 0)} + \biggr(1-\exp(-x/b)\biggr) 1_{(\lambda=0)},
$$

\Ni where $b>0$, $x\geq 0$ for $a\geq 0$. $0\leq x\leq -b/a$ for $a<0$. The parameter $\lambda$ and $b$ are called shape and scale parameters respectively. The link of the \textit{GPD} with \textit{UEVT} relates to the excess distribution over threshold $u$ for a cdf $F$ associated to a random variable $X$,  which is is defined by

$$
F_u(x)=\mathbb{P}(X-u\leq x | X>u)=\frac{F(x+u)-F(u)}{1-F(u)}, \ \ 0\leq x \leq uep(F)-u.
$$

\Bin The mathematical expectation of that conditional law is called the mean excess function is defined by

$$
e(u)-\mathbb{E}(X-u | x>u), \ u \in \mathbb{R}.
$$

\Bin We have the following important theorem

\begin{theorem}\label{pbh}

There exists a function $b(u)$ of $u\in \mathbb{R}$ such that

$$
\lim_{u\rightarrow uep(F)} \sup_{0\leq x\leq uep(F)-u} |F_u(x)-G_{\lambda,b(u)}(x)|=0
$$

\Bin if and only if $F \in G_{\lambda}$

\end{theorem}

\Bin It is known that for $0\leq \lambda<1$, the mean of $G_{\lambda,b}$ is $b/(1-\lambda)$. In the frame of Theorem \ref{pbh}

$$
e(u)\approx \frac{b}{1-\lambda} + \frac{\lambda}{1-\lambda} u
$$

\Bin for $u$ large enough. So the empirical methodology suggested for estimation a heavy tail consists in observing the empirical excess function from the data

$$
e_n(u)=\frac{\sum_{1\leq i \leq n} (X_i-u) 1_{X_i>u}}{\sum_{1\leq i \leq n}  1_{X_i>u}},
$$

\Bin and to check whether or not the curve $(u,e_n(u))$ is approximately a straight line for $u$ near \textit{uep(F)}. (see the paper by \cite{baMEF} for empirical estimation of the mean excess function by confidence bounds).\\

\Ni After this quick round up of the \textit{UEVT}, we are going to use some keys elements of it in our empirical studies.\\

\subsection{Scope of the study and organization of the paper} 

\noindent As outlined before, this research is concerned with the Chinese insurance market and
aims to identify the best distribution in modeling extreme claim sizes stock
returns by using the peak over threshold method (POT) to identify the best
distribution in modeling extreme claim sizes.\newline

\noindent The first classification is based on the gender and the second is
the experience of the driver.\newline

\noindent This new approach has many practical implications for both, the
insurer and the reinsurer. The main implication is that differentiating by
characteristics of the policyholders allows a fair premium paid by each
category of insured. Moreover, it allows an accurate estimation of the
distribution of extreme losses and a better determination of the limit for
individual claim size by the reinsurer in the case of the application of the
excess-of-loss reinsurance strategy.\newline

\noindent The remainder of this paper is organized as follows: The Section %
\ref{sec1} is a general introduction. The section \ref{sec2} describes the
data and sample statistics. The research methodology is discussed in Section %
\ref{sec3}. Section \ref{sec4} presents the empirical results. Section \ref%
{sec5} discusses the implications of the findings for the Chinese insurance
market. Summary and concluding remarks are given in Section \ref{sec6}.

\section{Data and sample statistics}\label{sec2}

\noindent In this empirical investigation, the dataset consists of a sample
of 405 177 observations for 4-wheeled vehicles and motorcycles from an
insurance company of the Hubei Province headquartered in Wuhan (the
insurance company covers the entire province of Hubei). The name of the
insurance company is omitted for security reasons. The data covers five
calendar years between 2012 and 2017. The data contains information about
the characteristic of the policyholders, the insured car and variables
related to the claims. In this paper, we mainly focus on the claim sizes,
the gender and the experience of the drivers.\newline

\noindent In this section, we particularly describe the claim sizes. We
proceed by a logarithm transformation to allow a better investigation of the
distribution of the claim amount. The logarithm transformation is also
recommended for the study of the extreme distribution. A complete
description of the portfolio is upon request in the supplementary documents.\\

\begin{figure}
	\centering
		\includegraphics[width=1.1\textwidth]{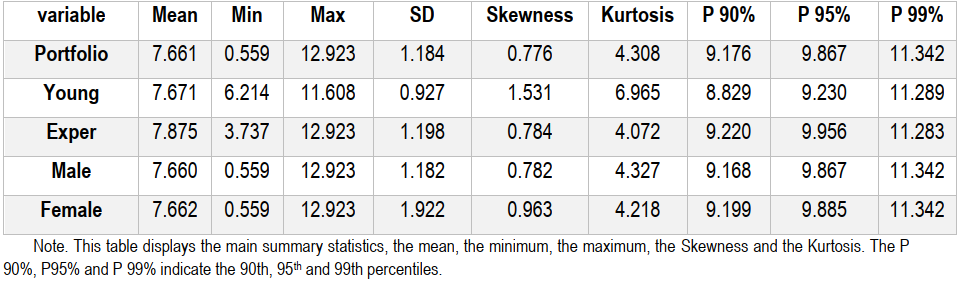}
		\caption{Summary statistics for the log claim sizes}
	\label{tab1}
\end{figure}

\noindent  \textbf{Note}. Table \ref{tab1}  displays the main summary statistics, the mean, the minimum, the maximum, the Skewness and the Kurtosis. The P 90\%, P 95\% and P 99\% indicate the $90^{th}$, $95^{th}$ and $99^{th}$  percentiles.\\

\noindent Table \ref{tab1} presents descriptive statistics of the claim sizes after logarithm transformation for the five samples. The mean of the claim sizes is likely to be equal for the five cases. The skewness value is positive for all datasets, indicating that distributions of the claims sizes are right skewed indication the presence of an upper tail. Moreover, Kurtosis exceeds the reference value of the Gaussian distribution (equal to 3) for all cases. These preliminary statistics corroborate the non-normality of the studied distributions. The examination of the different percentile shows large gaps between the calculated quantiles and the mean for all datasets. This gap may be considered as a serious problem for the insurer. Having estimated the pure premium using the mean of the claim sizes, the insurer may support additional losses related to the presence of these gaps. Accordingly, the insurer has to examine in more in-depth the claim sizes in the upper tail of the distributions.\\

\noindent The non-normality of the claim sizes justifies using other distributions characterized by the presence of a right tail as the skewness indicates.\\

\begin{figure}
	\centering
		\includegraphics[width=1\textwidth]{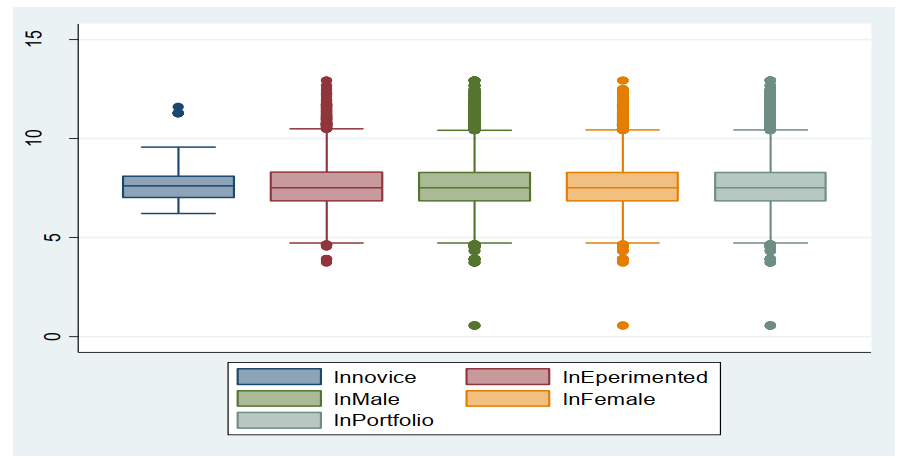}
		\caption{ The Box-plot of the claim sizes distributions}
	\label{fig1}
\end{figure}

\noindent Figure \ref{fig1} clearly shows the presence of extreme values for the entire portfolio as well as the different studied groups of insureds. The Boxplot helps only to show and detect the presence of the extreme values in a dataset. However, it did not allow for modeling the distribution of the detected extreme observations.

\section{Methodology} \label{sec3}

\noindent In this paper, we assume that the sequence $X_{1},X_{2},...,X_{n}$ of successive claim sizes consists of independent and identically distributed random variables generated by distribution $F_{X}$  of generic random variable $X$. The $n^{th}$ moment of the claim size distribution is usually denoted by $\mu _{X}^{\left( n\right) }=E\left( X^{n}\right) =\int_{0}^{\infty }$ $x^{n}$ $dF_{X}\left(
x\right) $. For $n=1$ we write $\mu _{X}=\mu _{X}^{\left( 1\right) }$. The variance of the claim sizes is written as $Var(X)=\mu_{X}^{\left( 2\right) }-\left( \mu _{X}\right) ^{2}$. The aggregate claim
amount at time $t$ can be written as $Y\left( t\right) =\sum_{i=1}^{N(t)} X_{t}$ and has the distribution function $F_{Y(t)}\left( y\right) =P(Y(t)\leq y)$. In any case, more specific information on the interdependencies within and between the two processes $N(t)$ and $X(t)$, describing the number and the size of the claims, is crucial.\\

\noindent In this paper we will focus only on the claim amount distribution. Thus, there is no impact of the dependence between the claim size and claim number on our modeling. However, this study will investigate the claim amounts, in particular, those reflecting the dangerous risks.

\subsection{ Concepts of risks and dangerous risks}

\noindent The risk, in its most general form, can be defined as uncertainty associated with a future outcome or event. To apply this more specifically to insurer activity, we can say that risk is the expected variance in losses or in the claim amounts. Statistically, as suggested by \cite{Rolski}, any non-negative random variable or its distribution is frequently called a risk. Therefore, any distribution, concentrated on the non-negative half-line, can be used as a claim size distribution. However, it will be worthwhile to make a distinction between "well-behaved" distributions and dangerous distributions with a heavy tail. The concepts of well-behaved or heavy-tailed distributions belong to
the common vocabulary of actuaries. \cite{Rolski} formalize these two concepts in a mathematically sound
definition. We also find an interest introduction to extreme value see \cite{Lo}.\\

\noindent The class of well-behaved distributions consists of those distributions with an exponentially bounded tail. This condition means that large claim sizes are not impossible, however their occurrence probabilities decreases exponentially fast to zero as the fixed threshold becomes larger and larger. In contrast, for the heavy tailed distributions there is no proper exponential bound and huge or extreme claims are getting more likely. A natural nonparametric class of heavy-tailed claim size distributions is the class of sub-exponential distributions (e.g. lognormal, Pareto and Weibull distributions with shape parameter smaller than 1). For more details about the statistical proprieties of sub-exponential distributions see, \cite{Rolski}, \cite{Omey} or more recently \cite{Lu} how provide some asymptotic results of the ruin probabilities in renewal risk model with some strongly sub-exponential claims and they obtain the asymptotic upper and lower bounds the studied distributions.\\

\noindent The sub exponential distributions are used to detect and predict the large claim amounts. However, a number of sub-exponential distributions are not defined in the domain of attraction of an extreme value distribution, \cite{Goldie}.\\

\noindent Since we are interested in modeling the claim sizes in the tail, we will focus only on the distributions defined in the domain of attraction of an extreme value distribution.

\subsection{Modeling large claim amounts}

\noindent It is obvious that the detection of large or extreme claim size distributions is one of the main concerns of the practicing actuary in insurance and reinsurance companies. The Extreme Value Theory is a tool for estimating the tails of a distribution. Two types of extreme value theory are used to this purpose, the classical extreme value theory (EVT) and the peak over threshold method (POT). In this paper, we will focus on the second method, which is suitable to our purposes. The next section gives a brief overview of the used extreme value methodology; first we will introduce the concepts of VaR and the expected shortfall. Then we will describe the peak over threshold method.

\subsubsection{Value-at-Risk and Expected Shortfall}

\noindent A very common risk measure in the financial world is the Value-at-Risk ($VaR$). This is in fact nothing else but at 95\%. For a distribution function $F$, $VaR$ is the $q^{th}$ quantile of $F$: $VaR$ $_{q}=F^{-1}\left( q\right) $.\\

\noindent For some $q$, typically in the domain $q\in \left( 0,95;1\right) $ $VaR$ provides an upper bound for a loss that is only exceeded on a small proportion of occasions, sometimes referred to as a confidence level. $VaR$  has been criticized as a risk measure, since it is not necessarily subadditive: There are cases where a portfolio can be split into sub-portfolio such that the sum of the $VaR$ for the sub-portfolios is smaller than the $VaR$ for the total portfolio. Further, $VaR$ gives
no information of the potential size of the loss exceeding $VaR$.\\

 \noindent Therefore, it has been proposed to use the expected shortfall (ES) or " tail conditional expectation" instead of $VaR$. $ES$  is the expected size of a loss exceeding $VaR$ : $ES_{q}:=E\left[ X\text{ }|\text{ }X\text{ }>\text{ }VaR_{q}\right] $.

\subsubsection{Peaks-Over-Threshold (POT)}

\noindent The most common group of models are the Peaks-Over-Threshold (POT) models. These are models for all large observations that exceed a high threshold. The POT models are generally considered to be the most useful for practical applications. With the POT class of models, one may further distinguish two styles of analysis. There are semi-parametric models, built around the so called Hill estimator (and its relatives), and the fully parametric models, based on the generalized Pareto distribution (GPD).Both classes are theoretically justified and empirically useful when used correctly.\\

\noindent  \textbf{The distribution of exceedances}\\

\noindent Let $X_{1},...,X_{n}$ be identically distributed (not independent) random variables with unknown distribution function\ $F(x)=\mathbb{P}\left( X\leq x\right) $. Given a high threshold\ $u_{n}$, we index each observation exceeding $u_{n}$ and obtain another sample $\{Y_{1},...,Y_{N_{u }}\}$
\ \ $N_{u }\leq n.$\\

\noindent Consider the $i.i.d$ case. Each point has the same chance to exceed the threshold with success probability $\mathbb{P}\left( X_{i}>u_{n}\right) ,$ $i=1,...,n.$ Hence, the number of exceeding observations is:\\

\noindent $N_{u_{n}}:=\#\{i:X_{i}$ $>u_{n},$ $i=1,...,n\}=\sum\limits_{i=1}^{n}1_{X_{i}>u_{n}}$.\\

\noindent $\ N_{u_{n}}$ follows a binomial distribution with parameters \ $n$ and\ \ $\mathbb{P}\left( X_{i}>u_{n}\right) $. Now a limit process can be derived by letting the sample size $n$ tend to infinity and, simultaneously, increasing\ $u_{n\text{ }}$ in the correct proportion: If for some $\tau >0,n\mathbb{P}\left( X_{i}>u_{n}\right) \longrightarrow \tau >,n\longrightarrow \infty $. Then, by a classical theorem \bigskip $N_{u_{n}}\underrightarrow{d}$ $\mathbb{ P}_{o}\left( \tau \right) $, where $\mathbb{P}_{o}$:  Poisson's variable. If $X_{1},...,X_{n},$ $i=1,...,n$ come from an absolutely continuous distribution, a suitable series $u_{n}$ \ can be found for every \ $\tau>0$. (See \cite{Embrechts}, chapter 3). Indexing all points\ $\{i:X_{i}$ $>u_{n}, $ $i=1,...,n\}$ in the interval $\left[ 0,n\right] $, this interval will grow larger whereas the indexed points will become sparser and sparser as\ $u_{n}$ increases with $n.$\\

\noindent  \textbf{The distribution of exceedances}\\

\noindent We are not only interested in when and how often the exceedances occur, but also in how large the excess $X-u|X>u$ is. Consider the conditional CDF of the excess observations $X-u,$\\

\noindent $F_{u}\left( x\right) =\mathbb{P}\left( X-u|X>u\right) .$ Or in terms of the underlying $F$ as $F_{u}\left( x\right) =\frac{F(x+u)-F\left( u\right) }{1-F\left( u\right) }.$\\

\noindent An important result in EVT is that for a very large class of distributions, it can be shown that:\\

\noindent $\underset{u\longrightarrow \infty }{\lim }$ $\underset{x\geq 0}{Sup}$\ $%
\left\vert F_{u}\left( x\right) -G_{\xi ,\beta \left( u\right) }\left( x\right)\right\vert
=0,$ where $G_{\xi ,\beta }$ is the CDF of the Generalized Pareto Distribution (GPD) :\\

\noindent  $G_{\xi ,\beta }\left( x\right) =\left\{ 
\begin{array}{ll}
1-\left( 1+\xi x/\beta \right) ^{-1/\xi }, & \xi \neq 0 \\ 
1-e^{^{-x/\xi }}, & \xi =0%
\end{array}%
\right. $\\

\noindent If $\xi \geq 0,$  the support of this distribution is $[0,\infty )$, for $\xi <0,$ the support is a compact interval. This distribution is generalized in the sense that it subsumes other distributions under a common parametric form. $\xi $  is the important shape parameter. The case $\xi =0$
corresponds to the exponential distribution.The case $\xi <0$ is known as a Pareto II distribution. If $\xi >0,$ then \ $G_{\xi ,\beta }\left( x\right) $ is a reparameterized version of the ordinary Pareto distribution, that has a long history in actuarial
mathematics. This is because the GPD is heavy-tailed when $\xi >0$. Whereas a normal distribution has finite moments of all orders, a heavy-tailed distribution does not, as already mentioned, possess a complete set of moments. In the case of a GPD with  $\xi >0,$  it is found that $E\left( X^{k}\right) =\infty $ for $k\geq $ $1/\xi .$ When $\xi =1/2$, the GPD has an infinite second moment, variance. When $\xi =1/4$, the GPD has an infinite fourth moment. Empirically, as mentioned in the section Empirical
properties of financial time series, it is often found that our series have an infinite fourth moment. The normal distribution cannot model these phenomena, but the GPD can be used to capture exactly this behavior.\\

\noindent The summarise, the excess distribution converges to a GPD. We have not defined the very large class of distributions for which this is true, but for our purposes it is enough to say that this one holds for all the common parametric continuous distributions (such as normal, lognormal, $\chi^{2},t,F,$ gamma...). Hence, the GPD is the natural model for the unknown excess distribution.\\

\noindent \textbf{Parameter estimation}\\

\noindent Finally, the parameters $\xi $ and $\beta $  must be estimated. A standard method is Maximum Likelihood (ML), where the joint PDF is maximized. However, in practice, this might be numerically troublesome if the data set is small, and one cannot rely on the asymptotic optimality properties of the
ML-estimators. Recall, that only the excess fraction of the set is used, and the used data set depends of course on the choice of threshold $u$. In our case, we typically have a very large number of observations, which is considered as enough to estimate our parameters.\\

\noindent For the choice of threshold $u$, the mean excess functions a useful tool.\\

\begin{center}
$e\left( u\right) =E\left( X-u|X>u\right) $
\end{center}

\noindent It can be estimated by the empirical function: $e\left( u\right) =\frac{1}{\#\{i:X_{i}>u_{n},i=1,...,n\}}\sum\limits_{i=1}^{n}\left( X_{i}-u\right)^{+}.$\\

\noindent For fat tails, $e\left( u\right) $ tends to infinity. For the GPD with $\xi>0$ it can be shown that $e\left( u\right) $ is a linearly increasing function. Hence, a possible choice of $u$ is given by the value for which $e\left( u\right) $ is approximately linear. In practice, this often gives
values such that the GPD is a good model for, very roughly, half of the sample.\\

\noindent \textbf{Parameter estimation}\\

\noindent Now, these results can be used to estimate tails and quantiles. Denote the
tail of $F$ by $\overline{F}=1-F$ These yields \\

\noindent $\overline{F}_{u}\left( y\right) =\mathbb{P}\left\{
X-u>y|X>u\right\} =\frac{\overline{F}_{u}\left( u+y\right) }{\overline{F}\left( u\right) }$ or $\overline{F}_{u}\left( u+y\right) =\overline{F}_{u}\left( y\right) \overline{F}\left( u\right) ,y\geq 0.$\\

\noindent Hence, an estimator of the tail $\overline{F}_{u}\left( y\right) $ (for values greater than $u$) can be obtained by estimating the tails $\overline{F}_{u}\left( y\right) $ and $\overline{F}\left( u\right) .\overline{F}\left( u\right) $ can be estimated by its empirical counterpart.\\

\noindent $\overset{\_\ast }{F}\left( u\right) =\frac{1}{n}\sum\limits_{i=1}^{n}I\left( X_{i}>u\right) =N_{u}/n$ \ and $\overline{F}_{u}\left( y\right) $ by the GPD, where the scaling function $\beta(u)$ has to be taken into account. This gives  $\overset{\_\ast }{F}\left( y\right) \approx \left( 1+\xi ^{\ast
}y/\beta ^{\ast }\right) ^{-1/\xi ^{\ast }}.$\\

\noindent The two parameters $\xi $ and $\beta $  have to be estimated, which is (theoretically) best done using ML.\\

\noindent Now, for a given $u$ this gives the tail estimator:  $\overset{\_\ast }{F}\left( u+y\right) \approx \frac{1}{n}\sum\limits_{i=1}^{n}\left( 1+\xi ^{\ast }y/\beta ^{\ast }\right) ^{-1/\xi ^{\ast }}.$\\

\noindent  For a given $q\in \left( 0,1\right) $, this function can be inverted to give an estimator of the q-quantile:\\

\noindent   $x_{q}^{\ast }=VaR_{q}^{\ast }=u+\frac{\beta ^{\ast }}{\xi ^{\ast }}\left( \left( \frac{n}{N_{u}}\left( 1-q\right) ^{\xi ^{\ast }}\right) -1\right).$\\

\noindent Hence, for a given probability $q>F(u)$, $VaR$  estimate $VaR_{q}^{\ast }.$

\section{The empirical results} \label{sec4}

\subsection{Q-Q plot against the exponential distribution}

\noindent Before applying the extreme value theory to investigate the behavior of the right tailed distributions it will be worthwhile assuming that the data are exponentially distributed. A commonly used methodology is to plot a quantile to quantile graphic against an exponential distribution.\\

\begin{figure}
	\centering
		\includegraphics[width=1\textwidth]{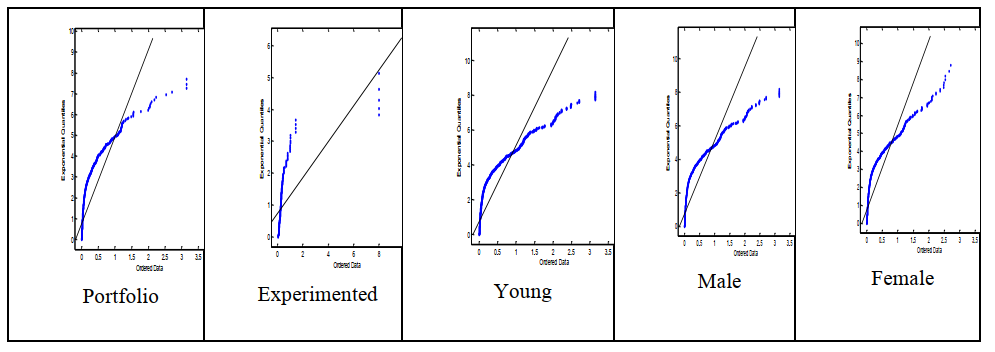}
		\caption{QQ-plot of the claim sizes data against standard exponential quantiles}
	\label{fig2}
\end{figure}

\noindent The five plots presented in figure \ref{fig2} show that a concave departure from the straight line in the QQ-plot. This behavior indicates a heavy tailed distribution of the claim sizes for the studied claim sizes series. The QQ-plot for a positive value of $\xi$ is linear; we can deduce that the adequacy of the data to the Generalized Pareto law seems to agree. A small percentage of insured persons cause responsible accidents with very high costs that the insurance company must bear and to which the insurance company applies special treatment.

\subsection{Peak over threshold results}

\noindent In this section, the estimation results of the GPD following the peak over threshold approach are presented. This investigation was conducted in order to compare the extremal behavior of the claim losses for the different studied groups of policyholders.\\

\subsubsection{ Threshold choice}

\noindent As a first step for estimating the GPD, a threshold needs to be selected from the maximum insurance claim sizes. To select the optimal threshold an assessment of mean residual life plot a threshold Choice plot and a L-moment plot were carried out following \cite{Coles}, \cite{Fersi} \cite{Farah}.\\

\noindent \textbf{Mean residual life plot:} To represent the mean residual life plot we use the theoretical mean of the GDP. For $X$ a random variable distributed as Generalized Pareto with parameters $\mu $, $\sigma $ and $\xi $. Theoretically we have :\\

\noindent $E\left[ X\right] =\mu +\frac{\sigma }{1-\xi }$ for $\xi <1.$\\

\noindent Empirically, if $X$ represents excess over a threshold $\mu _{0}$, and if the approximation by a GDP is good enough, we obtain:\ $E\left[ X-\mu_{0}|X>\mu _{0}\right] =\frac{\sigma _{\mu _{0}}}{1-\xi }.$ For all updated threshold $\mu _{1}$  such as $\mu _{1}>\mu _{0}$, excesses above the new threshold are also approximate by GPD after new parametrization. Thus, \\

\noindent $E\left[ X-\mu _{1}|X>\mu _{1}\right] =\frac{\sigma _{\mu _{1}}}{1-\xi }=\frac{\sigma _{\mu 0}-\xi \mu _{1}}{1-\xi }.$\\

\noindent The quantity $E\left[ X-\mu _{1}|X>\mu _{1}\right] $ is linear in $\mu _{1}$ and corresponds to the mean of excesses above the threshold $\mu _{1}$ which can easily be estimated using the empirical mean. Finally, a mean residual life plot consists in representing points:\\

\noindent $\left\{ \left( \mu ,\frac{1}{n_{\mu }}\sum\limits_{i=1}^{n_{\mu}}x_{i},n_{\mu }-\mu \right) \mu \leq x_{\max }\right\}.$\\

\noindent Where $n_{\mu }$ is the number of observations $x$ above the threshold $\mu,x_{i,n_{\mu }}$ is the $i^{th}$ observation above the threshold $\mu $ et $x_{\max }$ is the maximum of the observations $x$. Graphically, a threshold has to be selected when the mean residual plot is practically linear and the modified scale and shape estimates become constant.\\

\begin{figure}
	\centering
		\includegraphics[width=1\textwidth]{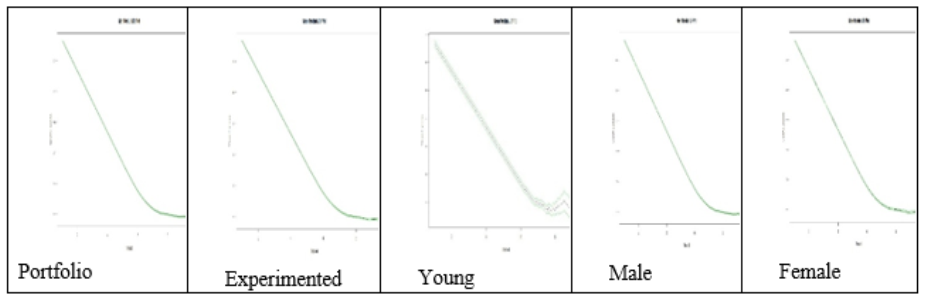}
		\caption{The mean residual life plot for the portfolio and the different insurance classes}
	\label{fig3}
\end{figure}

\noindent The figure \ref{fig3} indicate that the mean residual life plot of the maximum loss sizes thresholds, for the entire portfolio, is linear starting from a threshold of 8, where the line becomes more stable until about 9.5. For the experienced insureds, the figure \ref{fig3} indicates that the mean residual life plot is linear starting from 7.9 to 9.8. For the young drivers the mean residual life plot is linear starting from a threshold 8 and becomes more stable until about 9.5. Finally, for the males and females insureds, figure \ref{fig3} indicate that the mean residual life plot is linear from thresholds, respectively, 7.8 and 7.4 and becomes more stable until about 9.8 for the both classes.\\

\noindent \textbf{Threshold Choice plot}\\

\noindent The threshold choice plot is constructed using a random variable $X$ distributed as Generalized Pareto with parameters $\sigma _{0}$ and  $\xi _{0}$. Let \  $\mu _{1}>\mu _{0}$. The random variable $X|X>\mu _{1}$ is also GDP with parameters $\sigma _{1}=\sigma _{0}+\xi _{0}\left( \mu_{1}-\mu _{0}\right) $ and $\xi _{1}=\xi _{0}.$ Let $\sigma _{\ast }=\sigma_{1}+\xi _{1}\mu _{1}$, with this updated parameterization  $\sigma _{\ast }$ is independent of $\mu _{1}.$ Thus, estimates of $\sigma _{\ast }$ and $\xi_{1}$ are constant for all $\mu _{1}>\mu _{0}$ is a suitable threshold for
the asymptotic approximation. Then, the threshold choice plots represent the points defined by: $\left\{ \left( \mu _{1},\sigma _{\ast }\right) :\mu_{1}\leq x_{\max }\right\} $  where $x_{\max }$  is the maximum of the observations $X$.\\

\begin{figure}
	\centering
	  \includegraphics[width=0.45\textwidth]{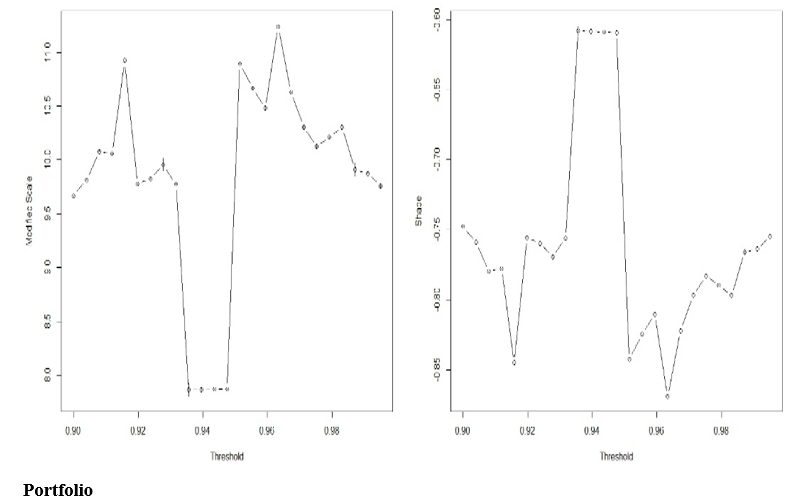}
		\includegraphics[width=0.45\textwidth]{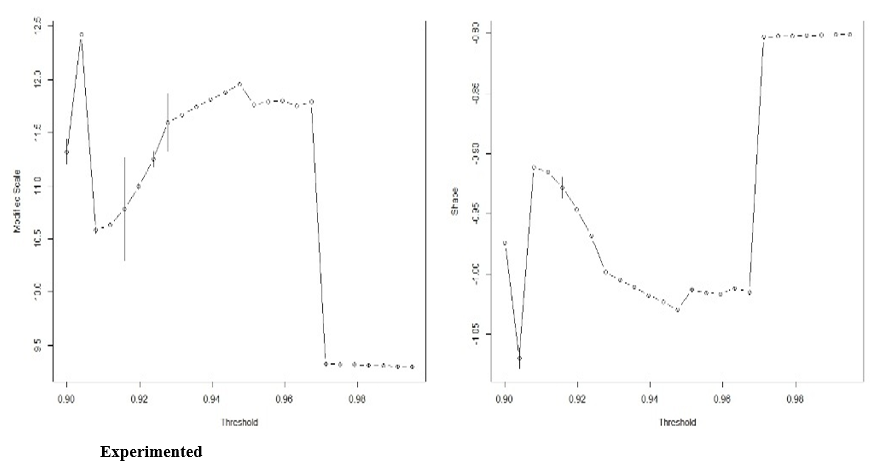}
		\includegraphics[width=0.45\textwidth]{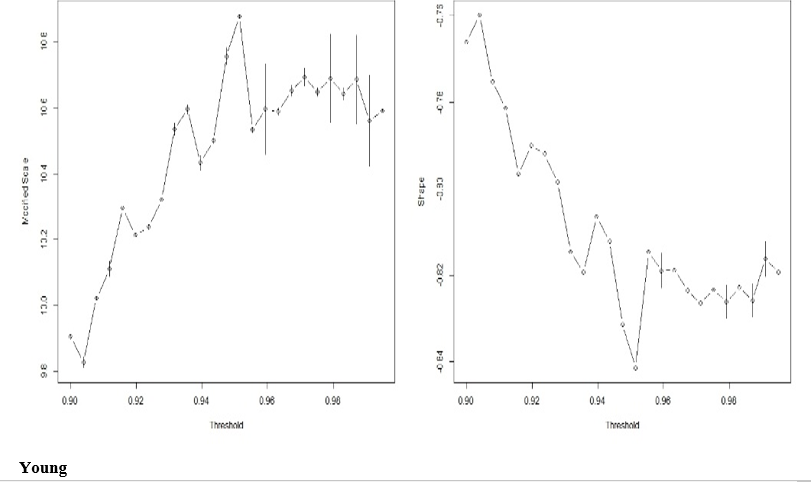}
		\includegraphics[width=0.45\textwidth]{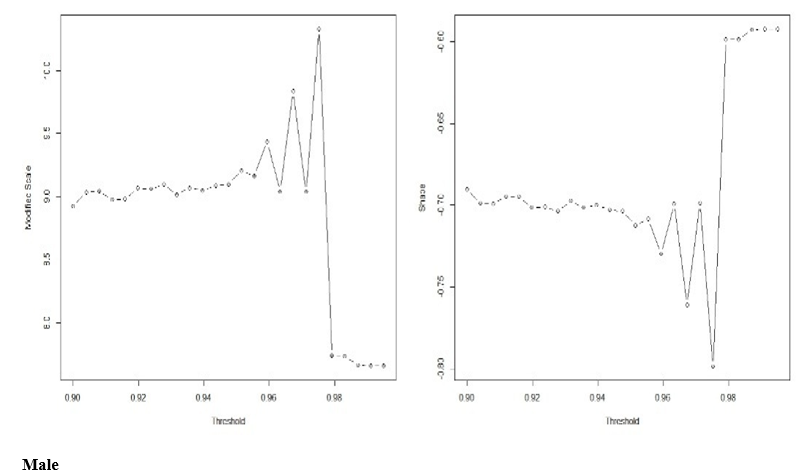}
		\includegraphics[width=0.45\textwidth]{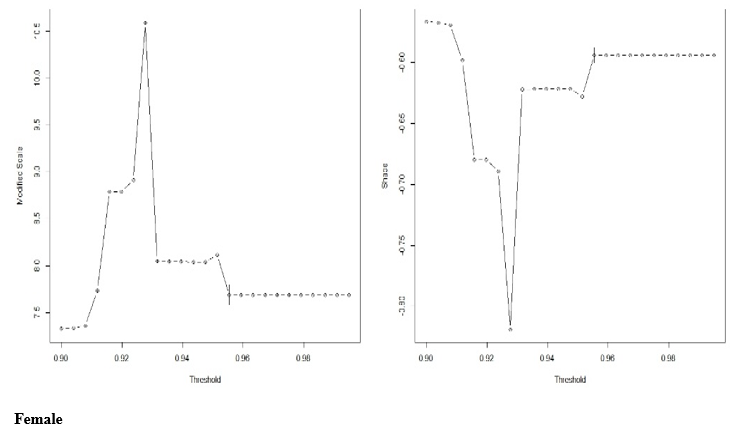}
		\caption{Results of the threshold choice plot function}
	\label{fig4}
\end{figure}

\noindent Figure \ref{fig4} displays results of the threshold choice plot function. We can see that thresholds can be contained in an interval with a minimum value of 90 and a maximum value of 0.98. For example, a threshold of 0.97 is a reasonable choice for males and a threshold of 0.95 is a reasonable choice for females. For the entire portfolio, a threshold of 0.95 would be a suitable choice. We must notice that in practice decision or threshold choices are not so clear-cut using the threshold choice plot function.\\

\noindent \textbf{L-moments plot:}\\

\noindent L-moments are the summary statistics for probability distributions and data samples. They provide measures of location, dispersion, skewness, kurtosis and other aspects of the shape of probability distributions or data sample. These measures are calculated from linear combinations of the ordered data values. For the Generalized Pareto distribution, the relation can be written as: $\tau _{4}=\tau _{3}\frac{1+5\tau _{3}}{5+\tau _{3}}$\\

\noindent where $\tau _{4}$  is the L-Kurtosis and $\tau _{3}$ is the L-Skewness. The L-Moment plot represents point defined by: $\left\{ \left( \widehat{\tau }_{3,u},\widehat{\tau }_{4,u}\right) :u\leq x_{\max }\right\}.$\\

\noindent Where $\widehat{\tau }_{3,u}$ and $\widehat{\tau }_{4,u}$ are estimations of the L-Kurtosis and L-Skewness based on excesses over threshold $u$ and $x_{\max }$ is the maximum of the observations $X$.\\

\begin{figure}
	\centering
		\includegraphics[width=1\textwidth]{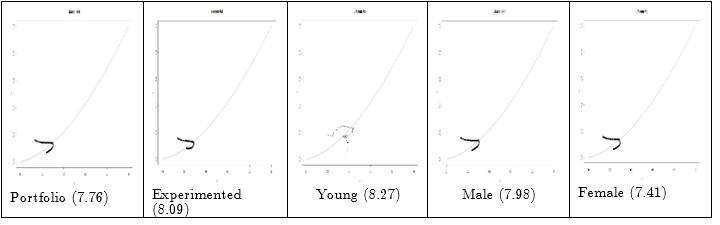}
		\caption{ Results of the L-moments plots}
	\label{fig5}
\end{figure}

\noindent The figure \ref{fig5} displays the L-Moment plots. The L-moments plots provide the value of the observations that will be considered as a threshold. The detected values are in log scales. For example, 7.76 is the threshold detect for the entire portfolio. For the experimented policyholders, the detected threshold is about eight.\\

\subsubsection{ Fitting the GPD}

\noindent After identifying a threshold above which the points are considered as extremes, we use a maximum likelihood method to estimate the two GPD parameters ($\xi$ and $\sigma $). In this section the estimation results of the GPD implementing the POT approach are presented. This indicates that the modeling of the distribution of the claim size by a law limited from the right is more adequate. Our results corroborate with those obtained by \cite{Pisarenko}.\\

\noindent For the different threshold, the estimated shape parameters are negative $\left( \xi <0\right) $, then the extreme claim sizes have a distribution on a bounded interval $\left[ 0,\beta /\xi \right] $. As highlighted previously, we select a threshold allowing for a stability of the estimated shape and scale parameter. The maximum stability is obtained for a threshold $u_{3}=10.5$. For various used thresholds, the two estimated parameters are statistically significant at $5\%$ significance level.
Since, the estimated shape parameter is stable and its value, for the different studied samples, $\widehat{\xi }>-0.5$  the estimators from ML are reliable \cite{Smith}.\\

\noindent The shape parameter estimates for the insurance claim sizes reveal some interesting facts. To compare between the different insurance groups it will be advisable to plot the different estimated shape parameters.\\

\begin{figure}
	\centering
		\includegraphics[width=1\textwidth]{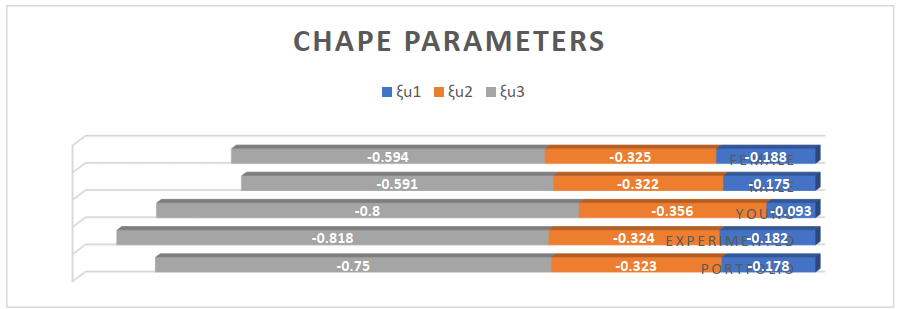}
		\caption{ The estimated shape parameters for the different studied insurance groups}
	\label{fig6}
\end{figure}

\noindent The shape parameter estimates for the insurance claim sizes reveal some interesting facts. To compare between the different insurance groups it will be advisable to plot the different estimated shape parameters.\\

\noindent  We recall that the larger the value of the shape parameter is, the larger the number of extreme events is. Hence, the tails of the claim size distribution become fatter when the shape parameter increases. In this prospect, the figure \ref{fig6} shows that the shape parameter for the experimented insureds is the smallest among all the remaining groups. This result is confirmed regardless of the selected threshold. Results also indicate that the class of experimented
insureds has lower number of extreme claim sizes than that for young policyholders. Moreover, results reveal that the number of extreme events are greater for male class than for female group of insureds.\\

\noindent Table \ref{tab2}  shows that the scale parameter is statistically significant at 5\% significance level for all the studied distributions. It is well known that the scale parameter is related to the volatility. Our results show that the volatility of the extreme claim sizes increases with the experimented and females policyholders. The lowest volatility is highlighted for the young drivers.\\

\noindent  Finally, results show that the interval estimates of the scale parameter are reliable since the range between the lower and the upper interval is very low for all the studied distributions.\\

\begin{table}
	\centering
		\includegraphics[width=1.1\textwidth]{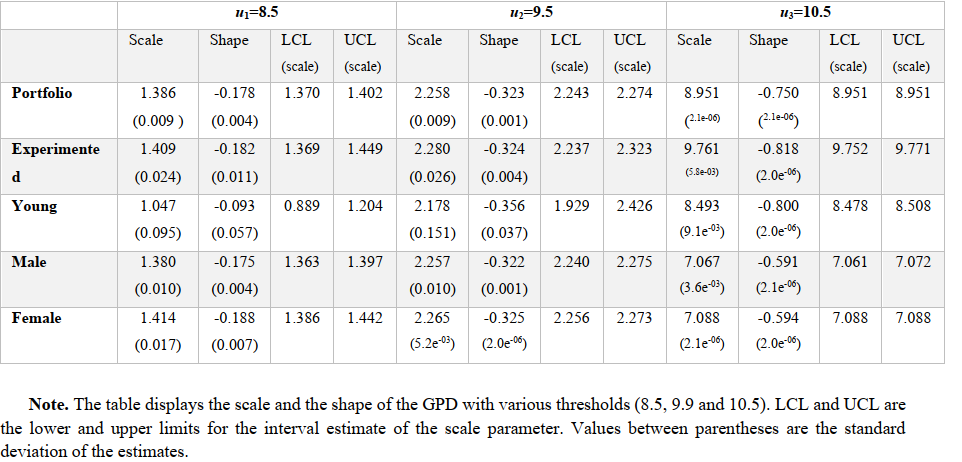}
		\caption{  Estimation results for the best model the POT approach (u1=8.5, u2=9.5 and u3=10.5)}
	\label{tab2}
\end{table}

\noindent  The comparison between the studied classes is essential for the insurer and actuary since it allows a better assessment of the pure premium that will be played by the insureds. The usual method of calculating the pure premium is to multiply the expected value of the loss amounts by the expected value of the number of accident. However, the expected value or the average is very sensitive to the presence of extreme values.\\

\noindent  Thus, the number and the amount of the detected extreme value can be used to adjust the pure premium by class of insureds.\\

\noindent Recall that the main objective of modeling the claim amounts is to estimate the coverage amount for claims. This means to determine a proper estimate of capital that allows insurers to be solvent in future years. Indeed, an insurance company needs to maintain a rational capital adequacy to avoid its insolvency probability. To address this need, the solvency 2 framework requires the use of Value at Risk ($VaR$) measure at a high level to enhance insured protection. To reach this aim, a useful application will be using the suitable distribution to calculate a value at risk by class and for the entire portfolio. Before calculating the $VaR$ we use deviance and the AIC criteria to select the suitable model for our datasets. The VaR will be calculated only for the entire portfolio as recommended by the capital requirement in solvency II.\\

\begin{table}
	\centering
		\includegraphics[width=1.1\textwidth]{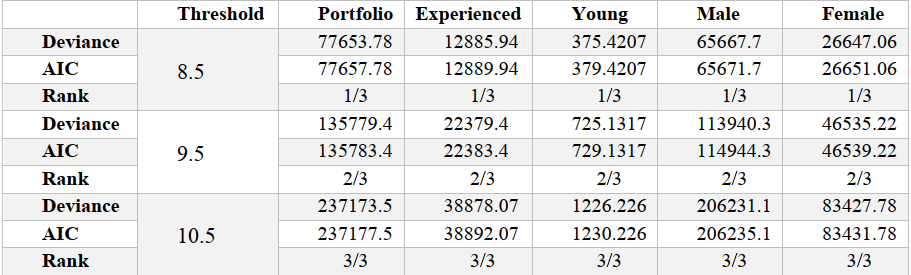}
		\caption{Selecting the best model}
	\label{tab3}
\end{table}

\noindent The results reported in table \ref{tab3} show that for all the studied samples the suitable model is a GPD with a threshold equals to 8.5.\\

\noindent In the insurance sector, the main element of sinistrality is described in the database by the amounts of compensation of responsible accidents declared by the insured. As announced in the previous chapters of our research, the skewness and the kurtosis are two very important parameters in modeling the tail of the distribution.\\

\subsubsection{Model diagnostic for the portfolio data}

\noindent After fitting extreme value model to data, next is to assess and interpret the fitted model based on the quantile and return levels computed using the inverse of the distribution function. We calculate and graphically present the return levels the annual scale, so that the 1-year return level is the level that is expected to be exceeded once in every one year. These parameters give example of very valuable information on the tail of distribution.\\

\begin{figure}
	\centering
		\includegraphics[width=1\textwidth]{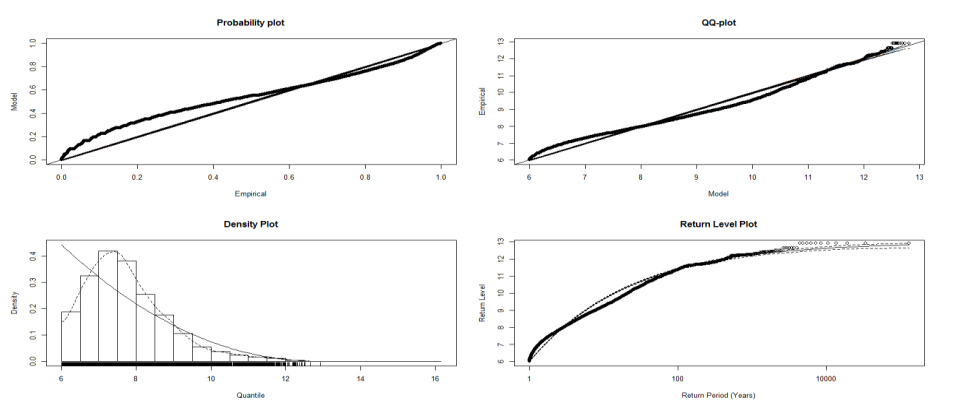}
		\caption{ Diagnostic plots of the fitted generalized Pareto distribution, threshold=8.5, for the portfolio.}
	\label{fig7}
\end{figure}

\noindent We use the function implemented in the software $R$; the result of function gives the name of the estimator, if a varying threshold was used, the threshold value, the number and the proportion of observations above the threshold, parameter estimates, standard error estimates and type, the asymptotic variance-covariance matrix and convergence diagnostic. We assess the adequacy and the validity of the fitted generalized Pareto distribution by analyzing the diagnostic plots presented in figure \ref{fig7}. Both the probability plot and the quantile plot indicate a reasonable extreme value fit because, the probability plot is almost linear and the quantile plot is practically linear.\\

\noindent  The return level plot highlights a slight convexity. Moreover, the return level plot indicates that the extreme values are within the 95\% confidence limits and the density plot adequately fit the upper right tail of the distribution. We finally conclude that the diagnostic plots do not raise any problem on the adequacy and the validity of the generalized Pareto fitting.\\

\noindent We also performed the diagnostic to the remaining studied datasets; the results are similar to those obtained for the whole portfolio. These results have important implications. First, having a distribution that fits well the large claim sizes will help the insurer in predicting the possible extreme event and the timing of its occurrence. Secondly, the accurate estimates of the scale and the shape of the large claim sizes will help in calculation of the value at risk and thus a correct estimation of the rational capital adequacy required by the Solvency 2.

\section{Discussion}\label{sec5}

\noindent In this paper, we highlighted the importance of the use of the Peak over threshold in detecting the extreme accident severities. Moreover, this approach helped in comparing between the probability of occurrence of normal and extreme accident severity of several driver's groups. 
The finding of the empirical assessment of the extreme accident severities that we  conducted are mostly stimulating and provides valuable information for road safety policy makers insurance companies as well.\\

\noindent First, our study contributes to develop strategies to improve china's road safety conditions. Indeed, it is well established that for the success of a road safety strategy, realistic quantified road safety targets should be set. In this prospect, our findings contribute to provide accurate quantitative measures that can be used to boost the road safety. Actually, the selection of suitable threshold, beyond which an accident is deliberated as extreme, provides a useful measure allowing separation between low-medium accident severity and extreme ones. This threshold can be considered as a bound between two clusters of accidents that have to be managed differently. Thus, road safety policymakers have to allocate resources and develop various appropriate safety plans according to the nature of severities (Low-medium or extreme). Specifically, the selected limit for the entire dataset exceeds 8.5, whereas the limit for young and experimented drivers are 8 and 7.9, respectively. The similar findings are stressed by gender with 8.5 for the entire dataset versus 7.8 for male and 8.5 the entire dataset versus 7.6 for female. For instance, resources and strategies engaged to improve road safety have to be allocated differently among young and experimented drivers taking into account the difference of probability of occurrence of extreme accidents.\\

\noindent Moreover, our finding provides an extreme value distribution by group of drivers. These distributions provide accurate prediction of extreme accident severity amounts. These predictions can be also exploited in outlining the strategy that should be engaged to reduce the number and the amounts of the extreme accidents.\\

\noindent Further, our results have important implications for the insurance market policymakers. Essentially, threshold estimation helps to identify the retention limits and achieve optimal reinsurance levels. The findings indicate that the retention limits vary among the studied groups and between the classes constructed based on different accident factors. More specifically, considering the experience in driving provides higher retention limit than that when allowing for gender as discrimination factor. Moreover, while the difference between subgroups is not significantly important, the difference between the retention limits of the entire insurance dataset and the different subgroups is suggestively high. Specifically, the selected retention limit of the entire dataset exceeds 8.5, whereas the retention limit for young and experimented drivers are 8 and 7.9, respectively. The similar findings are stressed by gender with 8.5 for the entire dataset versus $7.8$ for male and $8.5$ the entire dataset versus $7.6$ for female.\\ 

\noindent Overall, these results provide to the insurance company various strategies in selecting the appropriate retention levels and the reinsurance structure as well.  Specifically, changing the retention levels based on the entire data results or performing a discriminating factor (experience or age) will modify the risk profile and risk-based target capital for the insurer. 

\section{Conclusion}\label{sec6}

\noindent The aim of this work was to study the behavior of the extreme car insurance claim sizes by using the peaks-over-threshold method. A particular interest was given to the choice of the threshold to perform such analysis. Indeed, the choice of threshold affects the estimation of the GPD parameters. Our studies also aim to compare the extreme distribution for different risk classes. The following findings can be pointed out. There is no large gap in the selected threshold among the studied classes.


\begin{thebibliography}{99}

\bibitem[Karamata (1030)]{karamata} Karamata, J.(1930) Sur un mode de croissance r\'eguli\`ere des fonctions. \textit{Mathematica (Cluj)}, 4, 38-53 

\bibitem[Resnick (1987)]{resnick} Resnick, S.I. (1987). \textit{Extreme Values, Regular Variation and Point Processes}. Springer-Verbag, New-York. (MR0900810)

\bibitem[de Haan (1970)]{dehaan} de Haan, L. (1970). \textit{On regular variation and its
application to the weak convergence of sample extremes}. Mathematical Centre
Tracts, \textbf{32}, Amsterdam. (MR0286156)

\bibitem[Galambos (1985)]{galambos} Galambos, J. (1985). The Asymptotic theory of Extreme
Order Statistics. Wiley, Nex-York. (MR0489334)

\bibitem[de Haan and Feireira (2006)]{dehaan2} de Haan, L. and Feireira A. (2006). \textit{Extreme value
theory: An introduction. Springer}. (MR2234156)

\bibitem[Rolski \textit{et al.}(1999)]{Rolski} Rolski, T., Schmidt, V., and Teugels, J.(1986). \textit{Stochastic Processes for Insurance and Finance}. John Wiley \& Sons.

\bibitem[Omey(2006)]{Omey} Omey, A.M., 2006. Subexponential distribution functions. \textit{Journal of Mathematical Sciences}, 138(1), 5434-5449.

\bibitem[Lu and Bin Zhang(2016)]{Lu} Lu, D. and Bin Zhang, B., 2016. Some asymptotic results of the ruin probabilities in a two-dimensional renewal risk model with some strongly subexponential claims. \textit{Statistics \& Probability Letters}, 114, 20-29.

\bibitem[Goldie and  Resnick (1988)]{Goldie} Goldie, C.M. and  Resnick, S., 1988. Distributions that are both subexponential and in the domain of attraction of an extreme-value distribution. \textit{Advances in applied probability}, 20(4), 706-718.

\bibitem[Gnedenko (1943)]{gnedenko1943} Gnedenko, B.(1943) \textit{ Sur la distribution limite du terme maximum d'une série aléatoire}, Annals of Mathematics, 1943, 44, 423-453

\bibitem[CICR (2017)]{CICR} CICR.(2017)\textit{China insurance statistics report 2016}. Available at http://www.circ.gov.cn/web/site0/tab5257/2017.
	
\bibitem[Lozano-Perez (2012)]{Lozano-Perez} Lozano-Perez, T.(2012). \textit{Autonomous Robot Vehicles}. Springer Science \& Business Media.

\bibitem[Abraham \textit{et al.}(2016)]{Abraham} Abraham, H., Lee, C., and Brady, S.(2016). \textit{Autonomous Vehicles, Trust, and Driving Alternatives: A Survey of Consumer Preferences}. AgeLab, Massachusetts Institute of Technology.

\bibitem[Mao(2017)]{Abraham} Mao S.(2017). \textit{Vehicles import market 2016 in China Consumption Daily}. Beijing.

\bibitem[Xian and Chiang-Ku (2018)]{Xian} Xian, X. and Chiang-Ku, F.(2018). \textit{Autonomous vehicles, risk perceptions and insurance demand: An individual survey in China}. Elsevier.

\bibitem[Embrechts \textit{et al.} (1997)]{Embrechts} Embrechts, P., Kl\H{u}ppelberg, C., and Mikosch, T.(1997). \textit{Modelling Extremal Events for Insurance and Finance}. Berlin, Springer, 1997.

\bibitem[Coles (2001)]{Coles} Coles, S.G.(2001). \textit{An Introduction to Statistical Modeling of Extreme Values}.Springer Verlag, New York.

\bibitem[Fersi \textit{et al.}(2011)]{Fersi}  Fersi, K., Boukhetala, K., and Ammou, S.B.(1997). \textit{Strat\'egie optimale de r\'eduction de l'intervalle de confiance pour l'estimateur de la prime ajust\'ee. Application en assurance automobile}. Hal.

\bibitem[Farah and Azevedo, (2017)]{Farah} Farah, H. and Azevedo, C.L.(2017). Safety analysis of passing maneuvers using extreme value theory.\textit{IATSS Research}, 41, 12-21.

\bibitem[Pisarenko and Rodkin(2010)] {Pisarenko} Pisarenko, V.F. and Rodkin, M.V.(2010). \textit{Estimation of the Probability of Strongest Seismic Disasters Based on the Extreme Value Theory}. Springer LINK.


\bibitem[Smith, (1985)] {Smith} Smith, R.L.(1985). Extreme value analysis of environmental time series: An application to trend detection in ground-level ozone (with discussion). \textit{Statistical Science}, 4, 367-393.

\bibitem[Lo, (2017)] {Lo} Lo, G.S.(2017). \textit{Weak Convergence (IIA) - Functional and Random Aspects of the Univariate Extreme Value Theory}, Spas Textbooks Series. Arxiv :DOI : http://dx.doi.org/10.16929/srm/2016.0009.

\bibitem[Lo (1986)]{lo86} L\^o, G.S. (1986). \textit{Sur quelques estimateurs de l'Index d'une loi de Pareto : Estimateur de Hill, de S.Cs\"org\H{o}-
Deheuvels-Mason, de de Haan-Resnick et loi limites de sommes de valeurs extr\^emes pour une variable al\'eatoire dans le domaine d'attraction de
Gumbel}. Th\`ese de doctorat. Universit\'e Paris VI.

\bibitem[Gumbel (1955)]{gumbel} Gumbel, E.J. (1955) \textit{statistical estimation of the endurance limit},Tchnical report T-3A, Departement of Engineering, Columbia Univ. press.,New-York.

\bibitem[Fisher and Tippet (1928)]{fishertippet} Fisher, R. and Tippet, L. \textit{Limiting Forms of the Frequency Distribution of the Largest or Smallest Member of a Sample} Proceedings of the Cambridge Philosophical Society, 1928, 24, 180-190


\bibitem[Lo et al. (2018)]{ips-wciia-ang} Lo G.S., K. T. A. Ngom M. and Diallo M.(2018). 
Weak Convergence (IIA) - Functional and Random Aspects of the Univariate Extreme Value Theory.
Arxiv : 1810.01625


\bibitem[Lo\`eve (1997)]{loeve} Lo\`{e}ve,M.,(1997).\textit{Probability theory. Tome 1.} Springer-verlag, 4th Edition.


\bibitem[Feller (1968)]{feller2} Feller W.(1968) \textit{An introduction to Probability Theory and its Applications. Volume 2}. Third Editions. John Wiley \& Sons Inc., New-York. 

\bibitem[Beirlant \textit{et al.} (2004)]{bgt} Beirlant, J., Goegebeur, Y. Teugels, J.(2004). \textit{%
Statistics of Extremes Theory and Applications}. Wiley. (MR2108013)

\bibitem[Ba \textit{et al.} (2004)]{baMEF} Ba A.D.,Deme E.H, Seck C.T. and  Lo G.S.(2016). Consistency Bands for the Mean Excess Function and Application to Graphical Goodness-of-fit Test for Financial Data. \textit{Journal of Mathematical research}, Vol. 8, (1).  http://dx.doi.org/10.5539/jmr.v8n1p42, pp. 42-64
\end{thebibliography}
\end{document}